\documentstyle{article}
\begin{document}

\centerline{ \Large \bf On
 Hubbard-Stratonovich Transformations}
 \centerline{\Large \bf  over Hyperbolic
Domains.} \vskip 0.5cm \centerline{ \large \bf Yan V Fyodorov}

\vskip 0.3cm

{\it Dedicated to Prof. L. Sch\"{a}fer on the occasion of his
$60^{th}$ birthday.}

\vskip 0.3cm

\vskip 0.3cm
\begin{abstract}
We discuss and prove validity of  the Hubbard-Stratonovich (HS)
identities over hyperbolic domains which are used frequently in
the studies on disordered systems and random matrices. We also
introduce a counterpart of the HS identity arising in disordered
systems with "chiral" symmetry. Apart from this we outline a way
of deriving the nonlinear $\sigma$-model from the gauge-invariant
Wegner $k-$orbital model avoiding the use of the HS
transformations.
\end{abstract}

\section{Introduction}

A considerable progress achieved over the past two decades in
understanding statistical properties of a single electron motion
in disordered and chaotic systems\cite{Efetov,Mirlin} is mainly
based on the nonlinear $\sigma$-model description. This concept
was originally proposed in the context of disordered systems by
Wegner\cite{Wegner}, and further clarified and developed to a
working tool in the papers by Sch\"{a}fer and Wegner\cite{SW} and
Pruisken and Sch\"{a}fer\cite{PS}. The authors reduced the problem
of evaluating the disorder-averaged correlation functions of the
resolvents (Green's function) of a Hamiltonian containing disorder
to an effective deterministic field theory with a very peculiar
underlying non-compact symmetry, which they called "hyperbolic".
Original derivations\cite{SW} used a specific microscopic model,
the Wegner's gauge invariant $k-$orbital model \footnote{Readers
not familiar with the $k-$orbital model and its relation to
nonlinear $\sigma-$model may consult a recent
preprint\cite{rigor}, or Appendix A of the present paper for a
short introduction.}. It was however soon understood, that the
reduction is valid under much more general conditions. Loosely
speaking, the nonlinear $\sigma$-model adequately describes
physics of the one-electron motion for scales longer then the
so-called mean free path. In the gauge invariant model the mean
free path is effectively zero, which makes the derivation
particularly transparent.

In all the mentioned work the reduction of the microscopic
disordered models to the nonlinear $\sigma-$ model made use of the
so-called replica limit. The last essential ingredient - the idea
of supersymmetry, i.e. the use of both commuting and anticommuting
variables to avoid problematic replica method - was introduced in
the theory by Efetov \cite{Efetov}. The resulting theory was
successfully tested in an important non-perturbative limit of
"zero" spatial dimension, where its predictions were shown to be
identical to those following from the theory of large random
matrices. After that within few years the
 supersymmetric version of the nonlinear $\sigma-$
model was accepted as a standard tool in condensed matter physics,
and also proved to be very useful in other fields
ranging from the theory of chaotic
scattering\cite{VWZ,FySo} to Quantum Chromodynamics\cite{VW}.

Successful as it was in applications, the nonlinear $\sigma-$
model is derived from underlying disordered Hamiltonian  along a
mathematically subtle procedure. One of the standard non-trivial
ingredients of the derivation is the exploitation of the so-called
Hubbard-Stratonovich identity \footnote{The original paper
\cite{SW} does not use the name
 Hubbard-Stratonovich, but such a terminology
became standard after Efetov's work\cite{Efetov}.}:
\begin{equation}\label{HS}
C_N e^{-\frac{1}{2}\mbox{\small Tr}\hat{A}^2}= \int {\cal
D}\hat{R}\,\,e^{-\frac{1}{2}\mbox{Tr}\hat{R}^2-i\mbox{\small Tr}
\hat{R}\hat{A}}
\end{equation}
where both $\hat{R},\hat{A}$ are square $n\times n$ matrices with
entries $R_{ij}$ and $A_{ij}$, respectively, and $C_N$ is some
$A-$independent constant. When we take these matrices to be
complex Hermitian (or real symmetric), use the corresponding "flat
differentials" ${\cal D}\hat{R}\propto \prod_{i=1}^n dR_{ii}
\prod_{i<j} d\,{\mbox Im}[R_{ij}]d\,{\mbox Re}[R_{ij}]$ as volume
elements for Hermitian case, and ${\cal D}\hat{R}\propto
\prod_{i=1}^n dR_{ii} \prod_{i<j} dR_{ij}$ for real symmetric
case, and integrate in the right-hand side over all degrees of
freedom independently from minus to plus infinity, the integral
amounts to a product of $n^2$ (resp. $n(n+1)/2$) one-fold Gaussian
integrals, and the identity is completely trivial. The tricky
point is that in the problem under consideration such a simple
choice of the integration manifold of matrices $\hat{R}$ is
prohibited by some extra requirements which impose restriction on
the structure of the matrices $\hat{A}$ and $\hat{R}$. In
particular, for the whole theory to be
 well-defined one has to ensure second (linear
in $\hat{A}$) term in the exponent in the right-hand side to be
{\it purely imaginary}, by the very way one makes use of the HS
identities. With a non-Hermitian choice of the matrices
$\hat{A},\hat{R}$ it is already a rather nontrivial task. For an
informal discussion of this point in a simple example see
\cite{my95}; rigorous and comprehensive treatment can be found in
the recent review paper by Zirnbauer\cite{Martin}.

The difficulty was first encountered and successfully solved by
Sch\"{a}fer and Wegner. For the model which stems on the
microscopic level from disordered Hamiltonians with broken
time-reversal invarance (e.g. due to presence of a magnetic field)
they suggested the following choice of the integration domain:
\begin{equation}\label{schweg}
\hat{R}=\lambda \hat{T}\hat{T}^{\dagger}+i\hat{P}
\end{equation}
where $\hat{T}\in U(n_1,n_2)$ is in the pseudounitary group of
complex $n\times n$ matrices, $n=n_1+n_2$, with integer $n_1\ge
1,n_2\ge 1 $. The inverse for such matrices is given by
$\hat{T}^{-1}=\hat{L}\hat{T}^{\dagger}\hat{L}$, with
$\hat{L}=\mbox{diag}({\bf 1}_{n_1},-{\bf 1}_{n_2})$ where ${\bf
1}_{n}$ stands for the identity matrix of size $n$, and
$T^{\dagger}$ stands for the Hermitian conjugate of $\hat{T}$. The
matrices $\hat{P}$ are Hermitian block-diagonal:
$\hat{P}=\mbox{diag}(\hat{P}_{n_1},\hat{P}_{n_2})=\hat{P}^{\dagger}$,
and the parameter $\lambda>0$ is arbitrary. For the case of
Hamiltonians respecting the time-reversal invariance the structure
of the integration manifold is very much the same, with the
pseudoorthogonal group $O(n_1,n_2)$ replacing the pseudounitray
one.

It is the manifold of matrices $\hat{T}$ that encapsulates
non-compact ("hyperbolic") symmetry of the problem, and at the
next stage gives rise to the interacting Goldstone modes described
by the nonlinear $\sigma-$model. With such a nontrivial choice of
the integration manifold the indentity (\ref{HS}) loses its
transparency, and its verification is a separate, nontrivial task.
The idea of the proof \cite{SW} is to interprete the corresponding
integral as going over a high-dimensional contour deformation of
a simple Hermitian matrix. Accurate implementation of this
argument can be found in the mentioned review paper by Zirnbauer
\cite{Martin}.

Although the Sch\"afer-Wegner parametrisation (\ref{schweg}) of
the integration manifold is admissible it was never much in use,
virtually abandoned in favour of an alternative one due to
Pruisken and Sch\"{a}fer\cite{PS}:
\begin{equation}\label{prusch}
\hat{R}=\hat{T}^{-1}\hat{P}\hat{T},\quad {\cal D} R =d\mu_H(T)dP_1
dP_2 \Delta^2[\hat{P}]
\end{equation}
where $\hat{P}=\mbox{diag}(\hat{P}_{n_1},\hat{P}_{n_2})$, with
$\hat{P}_{n_1}$ and $\hat{P}_{n_2}$ being real
diagonal\footnote{The convergence of the integral (\ref{HS2}) in
the text below requires, in fact, infinitesemal imaginary parts to
be included in variables $\hat{P}$ as
$\hat{P}=\mbox{diag}(\hat{P}_{n_1}-i0^{+}{\bf
1}_n,\hat{P}_{n_2}+i0^{+}{\bf 1}_n)$. This shift is henceforth
assumed implicitly. Similar remark is applicable also to the
formula (\ref{HSch4})  in the "chiral" case.}, and $\hat{T}$ is
again as in Eq.(\ref{schweg}). The notation $d\mu_H(T)$ is used
here for the invariant (Haar's) measure on the pseudounitary
group, and for any diagonal $n\times n$ matrix
$\hat{B}=\mbox{diag}(b_1,b_2,\ldots,b_n)$ we use the notation
$\Delta[\hat{B}]=\prod_{i<j}(b_i-b_j)$ for the associated
Vandermonde determinant. The matrices $\hat{A}$ relevant in
present context are always of the form
$\hat{A}=\hat{A}_{+}\hat{L}$, where $\hat{A}_{+}\ge 0$ is
Hermitian positive semidefinite, and $\hat{L}$ is the "signature"
matrix entering the definition of the pseudounitary group
$U(n_1,n_2)$, see above.  The Pruisken-Sch\"{a}fer type of
parametrization for the integration manifold is used in the
majority of applications (in particular, in the supersymmetric
version of the theory, implicitly in \cite{Efetov} and explicitly
in\cite{VWZ} and subsequent papers). Nevertheless, it seems that
validity of the corresponding Hubbard-Stratonovich transformation
Eq.(\ref{HS}) was never properly checked. In fact, for many years
it was taken for granted\cite{my95} that a kind of "shift of the
contour" argument should be valid for such a parametrization as
well. The belief was however challenged by Zirnbauer (see
criticism in\cite{Martin0}, and recently in \cite{Martin}) who
revealed via a thorough analysis serious difficulties in
implementing this type of argument for Pruisken-Sch\"{a}fer
parametrization. This observation makes the situation uncertain
and calls for further investigations.

It is worth mentioning, that recently an alternative method of
treating the hyperbolic symmetry was introduced which avoids the
use of any variant of the Hubbard-Stratonovich
integral\cite{my2002}, see a very informative discussion in
\cite{Martin}. This method can be successfully applied to
gauge-invariant $k-$orbital model, and we outline the
corresponding derivation of the (bosonic) nonlinear $\sigma-$model
in the Appendix A of the present paper (see also \cite{rigor}).
Nevertheless, the Hubbard-Stratonovich transformations remains an
important (frequently, the only available) tool for majority of
microscopic models, in particular for the popular model of band
matrices, see \cite{band} and discussions in \cite{Martin}, as
well as for the case of random potential ("diagonal") disorder.

The main goal of this paper is to provide a way of independent
verification of the validity of the Hubbard-Stratonovich identity
for the Pruisken-Sch\"{a}fer type of parametrizations. We succeed
in solving the problem in full generality for the pseudounitary
case. As to the pseudoorthogonal case, the situation turns out to
be more complicated and is in fact quite interesting. By
considering explicitly the simplest nontrivial case $n_1=n_2=1$ we
will show that to save the identity one has to discard the
properties for the factors entering the elementary integration
volume to be positive and for the constant $C_N$ to be real.
Finally, we will discuss and prove the counterpart of the
Hubbard-Stratonovich identity which arises naturally in the
studies on disordered systems with special ("chiral") symmetry,
and is also useful when investigating non-Hermitian random
Hamiltonians. Some technical details are provided in the
Appendices B and C.

\section{ Hubbard-Stratonovich identities over Pruisken-Sch\"{a}fer
domains.}
\subsection{Pseudounitary case.}
We start with proving (\ref{HS}) for the general case of the
pseudounitary Pruisken-Sch\"{a}fer domain
$\hat{R}=\hat{T}^{-1}\hat{P}\hat{T}$ parametrized by matrices
$\hat{T}\in U(n_1,n_2)$ and $\hat{P}=\mbox{diag}
(\hat{P}_1,\hat{P}_2)$ as in (\ref{prusch}) . It is easy to show
(see e.g. \cite{my2002}, Appendix A) that for strictly positive
definite matrices $\hat{A}_{+}>0$ the corresponding matrices
$\hat{A}=\hat{A}_{+}\hat{L}$ always can be parametrised as
$\hat{A}=\hat{T_A}\hat{\Lambda}\hat{T}_A^{-1}$, where
$\hat{\Lambda}=\mbox{diag}(\hat{\Lambda}_{n_1},\hat{\Lambda}_{n_2})$
is diagonal such that
$\hat{\Lambda}_{n_1}>0,\hat{\Lambda}_{n_2}<0$, and $\hat{T}_A$ is
a pseudounitary matrix from $U(n_1,n_2)$. Starting with this form
of the matrix $\hat{A}$, we see that the right-hand side of
(\ref{HS}) can be written as
\begin{equation}\label{HS1}
I^{(pu)}_{HS}(\hat{A})=\int
d\hat{P}\Delta^2[\hat{P}]e^{-\frac{1}{2}\mbox{Tr}\hat{P}^2} \int
d\mu_{H}(\hat{T})\exp\{-i\mbox{\small Tr}
\hat{T}^{-1}\hat{P}\hat{T}\hat{T}_A\hat{\Lambda}\hat{T}_A^{-1} \}
\end{equation}
Now it is obvious that the integral is independent of the matrix
$\hat{T}_A$. Indeed, that matrix can be absorbed into $\hat{T}$ by
the change of variables $\hat{T}\hat{T}_A\to \hat{T}$, exploiting
the cyclic invariance under the trace and the invariance property
of the Haar's measure: $d\mu_{H}(T)\equiv
d\mu_{H}(\hat{T}\hat{T}_A)$. Second observation is that due to
(block-)diagonal structure of the matrices $\hat{P},\hat{\lambda}$
the combination in the exponent
$\mbox{Tr}\left[\hat{T}^{-1}\hat{P}\hat{T}\hat{\Lambda}\right]$
stays invariant when $\hat{T}$ is multiplied with an arbitrary
unitary block-diagonal matrix of the form
$\mbox{diag}(\hat{V}_1,\hat{V}_2)$, with $\hat{V}_1\in U(n_1)$
 and $\hat{V}_2\in U(n_2)$. Thus,
the combination in question does not change if we replace
$\hat{T}$ with matrices $\hat{T}_0\in
\frac{U(n_1,n_2)}{U(n_1)\otimes U(n_2)}$ taken from the coset
space obtained by factorizing the original pseudounitary group
$U(n_1,n_2)$ by its maximal compact subgroup $U(n_1)\otimes
U(n_2)$. As the result, the integral - up to a multiplicative
constant - can be replaced by one going over the coset space
rather than the whole pseudounitary group. This is a very pleasing
fact, since such an integral has been recently shown\cite{fystra1}
to be exactly calculable with help of the so-called
Duistermaat-Heckman localization theorem, generalizing a similar
formula known for the unitary group\cite{IZHC}. The result of such
integration is given, again up to a multiplicative constant,
by\footnote{See the previous footnote on convergence of this
integral.}
\begin{equation}\label{HS2}
\int d\mu(\hat{T_0})e^{-i\mbox{Tr}
\left[\hat{T_0}^{-1}\hat{P}\hat{T_0}\hat{\Lambda}\right]} \propto
\frac{\det{\left[e^{-ip_{1i}\lambda_{1j}}\right]_{i,j=1}^{n_1}}
\det{\left[e^{-ip_{2i}\lambda_{2j}}\right]_{i,j=n_1+1}^n}}
{\Delta[\hat{P}]\Delta[\hat{\Lambda}]}
\end{equation}
Substituting (\ref{HS2}) back to (\ref{HS1}), we bring the latter
to the form
\begin{equation}\label{HS3}
I^{(pu)}_{HS}(\hat{A})\propto \frac{1}{\Delta[\hat{\Lambda}]} \int
d\hat{P}_1d\hat{P}_2\,
\Delta[\hat{P}]\,e^{-\frac{1}{2}\mbox{\small Tr}\hat{P}^2}
\det{\left[e^{-ip_{1i}\lambda_{1j}}\right]_{i,j=1}^{n_1}}
\det{\left[e^{-ip_{2i}\lambda_{2j}}\right]_{i,j=n_1+1}^n}
\end{equation}
Now we observe that
$\Delta[\hat{P}]=\Delta[\hat{P}_1]\Delta[\hat{P}_2]
\prod_{i=1}^{n_1}\prod_{j=n_1+1}^{n}(p_{1i}-p_{2j})$
and use the invariance of the integrand with respect to
any permutation of the indices of integration variables
in the set $p_{1i},\,\, i=1,\ldots,n_1$, as well as in the set
$p_{2j},\,\, j=n_1+1,\ldots,n$. Such an invariance allows one to
replace each determinantal factor in the integrand
with only one ("diagonal") contribution,
multiplying the whole integral with the factor $n_1!n_2!$.
Disregarding the multiplicative factors, we write the resulting
integral as
\begin{equation}\label{HS4}
I^{(pu)}_{HS}(\hat{A})\propto \frac{1}{\Delta[\hat{\Lambda}]} \int
d\hat{P}_1d\hat{P}_2\,
\Delta[\hat{P}]\,e^{-\frac{1}{2}\mbox{\small Tr}\hat{P}^2}
e^{-i\sum_{i=1}^{n_1}p_{1i}\lambda_{1i}}
e^{-i\sum_{i=n_1+1}^{n}p_{2i}\lambda_{2i}}
\end{equation}
and further use the following well-known identity\footnote{ This
formula can be for example derived starting from (\ref{HS}) for
Hermitian matrices $\hat{R},\hat{A}$, diagonalising $\hat{R}$ by
unitary transformation, and performing the unitary group
integration by the Itzykson-Zuber-HarishChandra
formula\cite{IZHC}. See Appendix B for a proof of a similar
expression in "chiral" case.}
\[
\int d\hat{P} \Delta[\hat{P}]\,e^{-\frac{1}{2}\mbox{\small
Tr}\hat{P}^2 \pm i\mbox{\small Tr}[\hat{P}\hat{\Lambda}]}\propto
\Delta[\hat{\Lambda}] e^{-\frac{1}{2}\mbox{Tr}\hat{\Lambda}^2}
\]
valid for any real diagonal matrices $\hat{P},\hat{\Lambda}$, . In
view of $\mbox{Tr}\hat{\Lambda}^2\equiv \mbox{Tr}\hat{A}^2$ the
latter result shows that the original integral (\ref{HS1}) is
equal, up to a multiplicative $A-$independent constant to
$e^{-\frac{1}{2}\mbox{\small Tr}\hat{A}^2}$. We thus proved the
required Hubbard-Stratonovich identity for matrices
$\hat{A}=\hat{A}_{+}\hat{L}$, with $\hat{A}_{+}>0$.

What remains to be shown is how to incorporate the case of
positive semidefinite $\hat{A}_{+}\ge 0$ in the above scheme,
which is crucial for applications. The problem is that if the
matrix $\hat{A}_{+}$ has zero eigenvalues, the corresponding
matrix $\hat{A}=\hat{A}_{+}\hat{L}$ may be not $T-$
diagonalizable.\footnote{ The
simplest relevant example is $\hat{A} = \left(%
\begin{array}{cc}
1 & -1 \\
1 & -1%
\end{array}%
\right) \;, $ which squares to zero: $\hat{A}^2=0$.}
  To this end introduce a parameter $\varepsilon>0$ and consider the integral:
  \begin{equation}
    \int \mathcal{D}R \, \mathrm{e}^{- \frac{1}{2} \mbox{\small Tr}\,
    (\hat{R} + i \varepsilon \hat{L})^2 - i
  \, \mbox{\small Tr}\hat{A} \hat{R}}= \mathrm{e}^{\frac{1}{2}\varepsilon^2 \mathrm{Tr}{\bf 1}_n  }
  \int \mathcal{D}R \, \mathrm{e}^{- \frac{1}{2} \mathrm{Tr}\,
    \hat{R}^2- \mathrm{i}
    \mathrm{Tr}\,( \hat{A}+  \varepsilon \hat{L}) \hat{R}} \; \;.
\end{equation}

From $\hat{A}=\hat{A}_{+}\hat{L}$, where $\hat{A}_{+}\ge 0$
 immediately follows that
$\hat{A}_{\epsilon}\equiv \hat{A}+ \varepsilon
\hat{L}=(\hat{A}_{+}+\varepsilon {\bf 1}_n)\hat{L}\equiv
\hat{A}_{\epsilon,+}\hat{L}$, where $\hat{A}_{\epsilon, +}$ is
already positive definite:
$\hat{A}_{\epsilon,+}=\hat{A}_{+}+\varepsilon {\bf 1}_n>0$. But
such matrices $A_{\epsilon}$ are always T-diagonalizable, and  the
above-given proof of the Hubbard-Stratonovich identity retains its
validity. Therefore, the integral is calculated as
$$
     \mathrm{e}^{\frac{1}{2}\varepsilon^2 \mathrm{Tr}{\bf 1}_n }
  \int \mathcal{D}R \, \mathrm{e}^{- \frac{1}{2} \mathrm{Tr}\,
    \hat{R}^2- \mathrm{i}
    \mathrm{Tr}\,(\hat{A}_{\epsilon} \hat{R})} =
 \mathrm{e}^{\frac{1}{2}\varepsilon^2 \mathrm{Tr}{\bf 1}_n  }
\mathrm{e}^{-\frac{1}{2}\mathrm{Tr}\,( \hat{A}_{\varepsilon})^2}=
\mathrm{e}^{-\varepsilon \mathrm{Tr}\, \hat{L}\hat{A}}
    \mathrm{e}^{-\frac{1}{2} \mathrm{Tr}\, \hat{A}^2}
$$
resulting in the $\epsilon-$modified version of the
Hubbard-Stratonovich identity:
\begin{equation}\label{HSeps}
 \int \mathcal{D}R \, \mathrm{e}^{- \frac{1}{2} \mathrm{Tr}\,
    (\hat{R} + \mathrm{i} \varepsilon \hat{L})^2 - \mathrm{i}
    \mathrm{Tr}\, \hat{A} \hat{R}}=\mathrm{e}^{-\varepsilon \mathrm{Tr}\, \hat{L}\hat{A}}
    \mathrm{e}^{-\frac{1}{2} \mathrm{Tr}\, \hat{A}^2}
\end{equation}
For $\varepsilon > 0$ this identity holds uniformly in $\hat{A}$,
including the case of non-diagonalizable matrices $\hat{A}$.

\subsection{Pseudoorthogonal case: $n_1=n_2=1$.}

The difficulty of proving (\ref{HS}) in the important case of
general pseudoorthogonal group $O(n_1,n_2)$ is due to lack of
integration formulas similar to Eq.(\ref{HS2}) for cosets
$\frac{O(n_1,n_2)}{O(n_1)\otimes O(n_2)}$. Under these
circumstances we will restrict ourselves by the first non-trivial
case $n_1=n_2=1$, which proves to be already very informative. The
matrix $\hat{P}$ in this case is $2\times 2$ diagonal:
$\hat{P}=\mbox{diag}(p_1,p_2)$, and the matrices $\hat{T}_0$ can
be explicitly parametrized in terms of the variable $\theta\in
(-\infty,\infty)$ as
$\hat{T}_0=\left(\begin{array}{cc}\cosh{\theta}&\sinh{\theta}
\\ \sinh{\theta}&\cosh{\theta}\end{array}\right)$. Then the $2\times 2$
 matrices $\hat{R}=\hat{T}_0^{-1}\hat{P}\hat{T}_0$ we are
 integrating over in Eq.(\ref{HS}) are explicitly given by:
 \begin{equation}\label{po1}
 \hat{R}=\left(\begin{array}{cc}
 \frac{p_1+p_2}{2}+\frac{p_1-p_2}{2}\cosh{2\theta}
 &\frac{p_1-p_2}{2}\sinh{2\theta}
\\ - \frac{p_1-p_2}{2}\sinh{2\theta}&\frac{p_1+p_2}{2}-\frac{p_1-p_2}{2}\cosh{2\theta}
\end{array}\right)
\end{equation}
 As we already mentioned, the matrices $\hat{A}$ must be of the
form $\hat{A}=\hat{A}_{+}\hat{L}$, where
$\hat{L}=\mbox{diag}(1,-1)$. We restrict ourselves in this section
only with  real symmetric matrices $\hat{A}_{+}>0$ for simplicity,
i.e. $\hat{A}_{+}=\left(\begin{array}{cc}a_{1}& a
\\ a & a_{2}\end{array}\right)>0$. Hence
$\hat{A}=\left(\begin{array}{cc}a_{1}& -a
\\ a & -a_{2}\end{array}\right)$, with the constraints
\begin{equation}\label{a}
a_{1}>0,\,a_{2}>0,\, |a|<\sqrt{a_{1}a_{2}}
\end{equation}

Naively, one may expect the choice of the volume element on such a
manifold in the form $d\hat{R}=|p_1-p_2|dp_1\,dp_2\,d\theta$ to be
"natural". We will however see below that taking such a choice we
end up in a trouble: the corresponding Hubbard-Stratonovich
formula (\ref{HS}) does not hold its validity any longer. Instead,
the correct choice of the "volume element" in our case turns out
to be:
\begin{equation}\label{vol}
d\tilde{R}=(p_1-p_2)dp_1\,dp_2\,d\theta
\end{equation}
This expression is not sign-definite any longer, and changes sign
at the line $p_1=p_2$; we will comment on this point shortly later
on.

After having specified all ingredients of the right-hand side in
(\ref{HS}) we can write down the corresponding integral explicitly
as
\begin{eqnarray}\label{po2}
&&I^{(po)}_{HS}=\int_{-\infty}^{\infty}dp_1\int_{-\infty}^{\infty}dp_2
(p_1-p_2) e^{-\frac{1}{2}(p_1^2+p_2^2)-i\frac{1}{2}(p_1+p_2)
(a_{1}-a_2)} \\ && \nonumber \times \int_{-\infty}^{\infty}d\theta
e^{-i\alpha\cosh{2\theta}-i\beta\sinh{2\theta}}
\end{eqnarray}
where we introduced the shorthand notations
$\alpha=\frac{1}{2}(a_1+a_2)(p_1-p_2),\, \beta=a(p_1-p_2)$. We
note that in view of (\ref{a}) holds
$\beta/\alpha=\frac{a}{\frac{a_1+a_2}{2}}\le \frac{a}{\sqrt{a_1
a_2}}<1$. Therefore, we can parametrize $\beta=u\sinh{\psi},\,
\alpha=u\cosh{\psi}$, where $u,\psi$ are real parameters. Then the
combination entering second exponent in (\ref{po2}) can be
rewritten as $\alpha\cosh{2\theta}+\beta\sinh{2\theta}\equiv
u\cosh{(2\theta+\psi)}$. Finally, introducing $\mu=2\theta+\psi$
as integration variable, we see that the integral over $\theta$
can be explicitly calculated as \cite{RG}(a):
\begin{equation}\label{po3}
\frac{1}{2}\int_{-\infty}^{\infty}d\mu e^{-iu\cosh{\mu}}=
-\frac{\pi}{2}\left[Y_0(|u|)+i\,\mbox{sgn}(u)J_0(|u|)\right]\equiv
K_0(iu)
\end{equation}
where $\mbox{sgn}(u)=\pm 1$ depending on the sign of the variable
$u$, and $J_0(z),Y_0(z),K_0(z)$ are Bessel, Neumann and Macdonald
functions of zero order, respectively\cite{RG}. In our case:
\begin{equation}\label{u}
u = (p_1 - p_2) s_a, \quad s_a\equiv
\sqrt{\left(\frac{a_1+a_2}{2}\right)^2-a^2}
\end{equation}
Substituting the result of integration back to (\ref{po2}) and
changing to integration variables $p_{\pm}=(p_1\pm p_2)$, one can
easily perform the Gaussian integral over $p_{+}$ and obtain
\begin{eqnarray}\label{po4}
I^{(po)}_{HS}=-i\pi^{3/2} e^{-\frac{1}{4}(a_{1}-a_2)^2}
\int_{-\infty}^{\infty}dp_{-}p_{-} e^{-\frac{1}{4}p_{-}^2}
\left[Y_0\left(|p_{-}|s_a\right)+i\,
\mbox{sgn}(p_{-})J_0\left(|p_{-}|s_a\right)\right]
\end{eqnarray}
This is exactly the point where we can most clearly see the
necessity of omitting the absolute value sign in the measure
(\ref{vol}). Indeed, had we maintained the modulus $|p_{-}|$ in
the above integral, the second term in the integrand, being odd in
$p_{-}$, would vanish and the remaining integral would be that
containing the Neumann function. Although it is well-defined, it
could not produce the structure necessary for the validity of the
identity Eq.(\ref{HS}). In contrast, when the factor $p_{-}$ in
the measure does not contain the absolute value, it is the first
term which vanishes, and the second term can be straightforwardly
integrated by using the identity\cite{RG}(b):
\[
\int_{0}^{\infty}dp \, p\,  e^{-b p^2}J_0(p \, c)=\frac{1}{2b}\,
e^{-\frac{c^2}{4b}}
\]
yielding, with $b\equiv1/4, \, c\equiv s_a$
\begin{eqnarray}\label{po5}
&&I^{(po)}_{HS}=-4i\pi^{3/2}
e^{-\frac{1}{4}(a_{1}-a_2)^2-\frac{1}{4}(a_{1}+a_2)^2+a^2}\\
\nonumber && = -4i\pi^{3/2}e^{-\frac{1}{2}(a_{1}^2+a_2^2-2a^2)}
\equiv -4i\pi^{3/2}e^{-\frac{1}{2}\mbox{\small Tr} \hat{A}^2}
\end{eqnarray}
exactly as required by the Hubbard-Stratonovich identity
Eq.(\ref{HS}).

The considered example (and also one in the next section) makes it
clear that the only consistent way of ensuring the validity of the
HS transformation over hyperbolic domains is to require absence of
the absolute value of non-positively defined factors in the
elementary volume. From geometric point of view the integration
domain in hyperbolic case consists of several (in the simplest
case two) disconnected pieces. One may notice that the factor in
the elementary volume (\ref{vol}) changes sign precisely when
passing from one such piece to a different one, being
sign-constant within each piece. Persistence of this structure for
$n>1$, as well as finding a comprehensive proof valid for any $n$
in the pseudoorthogonal case remains an interesting open question
deserving further attention.

\subsection{ Chiral variant of the Hubbard-Stratonovich identity}
Last decade new symmetry classes of random Hamiltonians attracted
a lot of interest due to numerous applications in various branches
of physics, see \cite{Martin1} for discussion and basic
references. In particular, the class of Hamiltonians with chiral
symmetry is pertinent for analyzing properties of Dirac fermions
in random gauge field background, and found applications in
Quantum Chromodynamics \cite{VW}, as well as in condensed matter
theory, see e.g \cite{fystra2} for more references and discussion.
When reducing analysis of such systems to the relevant nonlinear
$\sigma-$model, one encounters the following variant of the
Hubbard-Stratonovich identity:
\begin{equation}\label{HSch}
I^{(ch)}_{HS}(\hat{A},\hat{B})= C_N e^{-\mbox{\small
Tr}[\hat{A}\hat{B}]}= \int
 {\cal D}\hat{R_1}{\cal D}\hat{R}_2\,\,
 e^{-\mbox{Tr}\hat{R}_1\hat{R}_2-i\mbox{\small Tr}
[\hat{R}_1\hat{A}+\hat{B}\hat{R}_2]}
\end{equation}
In the simplest case the two involved matrices are related as
$\hat{A}^{\dagger}=\hat{B}\in GL(n,{\mathcal{C}})$ , i.e $\hat{A}$
is an arbitrary complex matrix, and one can make a natural choice
of the integration domain $\hat{R}_1^{\dagger}=\hat{R_2}\in
GL(n,{\mathcal{C}})$, with elementwise "flat measure" on it. The
linear in $\hat{A}$ term in the exponent is then purely imaginary
as required, and the identity
\begin{equation}\label{HSchsimp}
I^{(ch)}_{HS}(\hat{A},\hat{A}^{\dagger})= C_N e^{-\mbox{\small
Tr}[\hat{A}\hat{A}^{\dagger}]}= \int
 {\cal D}\hat{R}{\cal D}\hat{R}^{\dagger}\,\,
 e^{-\mbox{Tr}\hat{R}^{\dagger}\hat{R}-i\mbox{\small Tr}
[\hat{R}^{\dagger}\hat{A}+\hat{A}^{\dagger}\hat{R}]}
\end{equation}
 follows from the integral in the
right-hand side being the standard Gaussian one.

In the applications, however, the case of two unrelated Hermitian
positive semidefinite matrices $\hat{A}^{\dagger}=\hat{A}\ge 0,\,
\hat{B}^{\dagger}=\hat{B}\ge 0$ emerges as well, and the extra
convergency arguments necessitate to make the choice for the
integration domain to be compatible with that property (see
below). Again, ensuring the pure imaginary nature of the exponent
and subsequent verification of the Hubbard-Stratonovich identity
(\ref{HSch}) for such a domain is a non-trivial task which was
not, to the best of author's knowledge, yet accomplished in full
generality.

Our strategy in this case will be informed both by our experience
with the pseudounitary and pseudoorthogonal cases. Again
restricting ourselves in this section with $\hat{A}>0,\, \hat{B}>
0$ we observe that any pair of Hermitian, positive definite
$n\times n$ matrices $\hat{A},\hat{B}$ can be parametrized as (see
\cite{fystra2} appendix A)
\begin{eqnarray}
&& \hat{A}=\hat{T}_A\hat{a}\hat{T}_A^{\dagger},\,
\hat{B}=\left[\hat{T}_A^{\dagger}\right]^{-1}\hat{a}\hat{T}_A^{-1}
\\ && \mbox{where}\quad  \hat{a}=\mbox{diag}(a_1,\ldots,
a_n)>0,\quad
 \hat{T}_A\in \frac{GL(n,{\mathcal{C}})}
 {U(1)\times\ldots \times U(1)},
 \end{eqnarray}
that is the matrix $\hat{T}_A$ is a general complex with real
positive diagonal entries. This suggest an idea to parametrize the
integration manifold as
\begin{equation}\label{par}
\hat{R}_{1}=\left[\hat{T}^{\dagger}\right]^{-1}\hat{P}\left[\hat{T}\right]^{-1}
\quad,\quad \hat{R}_{2}=\hat{T}\hat{P}\hat{T}^{\dagger}
\end{equation}
in terms of a  real diagonal matrix
$-\infty<\hat{P}=\mbox{diag}(p_1,\ldots, p_n)<\infty$ and a
general complex matrix  $\hat{T}\in {\sf{GL(n,{\mathcal{C}})}}$.
Guided by our previous experience, the volume element in the new
coordinates is chosen by:
\begin{equation}\label{measure}
{\cal D}\hat{R}_{1}{\cal D}\hat{R}_2\propto \prod_{l=1}^{n}p_l
dp_l\prod_{l<m}(p_l^2-p_m^2)^2d\mu_H(\hat{T},\hat{T}^{\dagger})
\end{equation}
where $d\mu_H(T,T^{\dagger})$ is the invariant Haar's measure on
the group $\sf{Gl(n,{\mathcal{C}})}$. Note that despite the domain
of integration being $-\infty < p_i<\infty,\,\, i=1,\ldots,n$ the
volume element contains factors $p_i$ rather than $|p_i|$ as one
may naively expect.

Substituting such a parametrization into the integral (\ref{HSch})
and using the cyclic invariance under the trace, we have
$\mbox{Tr}[\hat{R}_1\hat{R}_2]=\mbox{Tr}\hat{P}^2$ and the
$T$-dependent term in the exponent is given by
$$
\mbox{Tr}[\hat{R}_1\hat{A}+\hat{B}\hat{R}_2]=
\mbox{Tr}\hat{P}\left[\hat{T}^{\dagger}\hat{T}_A
\hat{a}\hat{T}_A^{\dagger}\hat{T}+\hat{T}^{-1}
\left(\hat{T}_A^{\dagger}\right)^{-1}
\hat{a}\hat{T}_A^{-1}\left(\hat{T}^{\dagger}\right)^{-1}\right]
$$
Now we change variables $\hat{T}_A^{\dagger}\hat{T}\to \tilde{T}$
and exploiting the invariance of the measure
$d\mu_H(T,T^{\dagger})=d\mu_H(\tilde{T},\tilde{T}^{\dagger})$
satisfy ourselves that we need to deal with the following group
integral:
\begin{equation}\label{HSch3}
\int_{T\in \sf{Gl(n,{\mathcal{C}})}} d\mu_H(T,T^{\dagger}) \exp{
-i\mbox{Tr}\hat{P}\left[\hat{T}^{\dagger}
\hat{a}\hat{T}+\hat{T}^{-1}\hat{a}\left(\hat{T}^{\dagger}\right)^{-1}\right]}
\end{equation}
Again, due to diagonal structure of the matrices $\hat{P},\hat{a}$
the integrand is not changed if we replace matrices $\hat{T}\in
\sf{Gl(n,{\mathcal{C}})}$ with $\hat{T_0}\in
\frac{GL(n,{\mathcal{C}})}{U(1)\times\ldots \times U(1)}$. This
fact provides us with the possibility of exploiting one more
integration formula discovered in \cite{fystra2}\footnote{Although
the formula itself is correct, its derivation in Appendix B of
\cite{fystra2} was not accurate enough. For this reason we include
the outline of the correct derivation in the Appendix C to this
paper.}:
\begin{equation}\label{HSch4}
\int_{T\in \sf{\frac{Gl(n,{\mathcal{C}})}{[U(1)\times\ldots \times
U(1)]}}} d\mu_0(T_0,T_0^{\dagger}) e^{
-i\mbox{Tr}\hat{P}\left[\hat{T}_0^{\dagger}
\hat{a}\hat{T}_0+\hat{T}_0^{-1}\hat{a}\left(\hat{T}_0^{\dagger}
\right)^{-1}\right]}\propto
\frac{\det{\left[K_0(2ip_{i}a_{j})\right]_{i,j=1}^{n}}}
{\Delta[\hat{P^2}]\Delta[\hat{a^2}]}
\end{equation}
where $d\mu_0(\hat{T}_0,\hat{T}_0^{\dagger}) =d\hat{T}_0
d\hat{T}_0^{\dagger}
\det{\left[\hat{T}_0\hat{T}_0^{\dagger}\right]}^{-n+\frac{1}{2}}$
is exactly the volume element on the coset space manifold.
Substituting this result back to the integral (\ref{HSch}),
cancelling one of the Vandermonde factors coming from the volume
element (\ref{measure}), and again exploiting the invariance
properties of the integrand with respect to permutation of
integration variables in the set $p_1,\ldots,p_n$ (cf. discussion
after (\ref{HS3})), and disregarding $A-$ independent
multiplicative factors, we bring the integral of (\ref{HSch}) to
the form
\begin{eqnarray}\label{HSch6}
&&I^{(ch)}_{HS}(\hat{A},\hat{B})\propto
\frac{1}{\Delta^2[\hat{a}^2]} \int_{-\infty}^{\infty}
\prod_{l=1}^n dp_l p_l\, \Delta[\hat{P}^2]\,e^{-\sum_{l=1}^n
p_l^2}\prod_{l=1}^n K_0(2ip_{l}a_{l})\\
&& \propto
\frac{1}{\Delta^2[\hat{a}^2]}\sum_{S_{\alpha}}(-1)^{S_{\alpha}}\prod_{l=1}^{n}\left[
\int_{-\infty}^{\infty} dp p\ e^{-p^2}p^{2 s_l} K_0(2ipa_l)\right]
\end{eqnarray}
where we used the expansion of the Vandermonde determinant as
$\Delta[\hat{P}^2]=\sum_{S_{\alpha}}(-1)^{S_{\alpha}}\prod_{l=1}^{n}p^{2
s_l}$ in terms of the sum over $n!$ permutations
$S_{\alpha}=(s_1,s_2,\ldots,s_n)$ of the index set
$(1,2,\ldots,n)$, with $(-1)^{S_{\alpha}}$ standing for the parity
of the permutation. Now we substitute the formula (\ref{po3}) for
the Bessel function of the imaginary argument to the above
expression, and see again that the term with Neumann function is
multiplied with the odd in $p$ factor and vanishes upon
integration, yielding
\begin{eqnarray}\label{HSch7}
&&I^{(ch)}_{HS}(\hat{A},\hat{B}) \propto
\frac{1}{\Delta^2[\hat{a}^2]}\sum_{S_{\alpha}}(-1)^{S_{\alpha}}\prod_{l=1}^{n}\left[
(-i\pi) \int_{0}^{\infty} dp p\ e^{-p^2}p^{2 s_l}
J_0(2pa_l)\right]
\end{eqnarray}
where we used $\mbox{sign}(a_i)=1$.  Finally we invert the
operation of the Vandermonde determinant expansion and show in the
Appendix B that the resulting integral can be evaluated as :
\begin{eqnarray}\label{HSch8}
\frac{1}{\Delta^2[\hat{a}^2]} \int_{0}^{\infty} \prod_{l=1}^n dp_l
p_l\, \Delta[\hat{P}^2]\,e^{-\sum_{l=1}^n p_l^2}\prod_{l=1}^n
J_0(2p_{l}a_{l})\propto e^{-\sum_{l=1}^n a_l^2}=e^{-\mbox{\small
Tr}\hat{A}\hat{B}}
\end{eqnarray}
exactly as required by the Hubbard-Stratonovich identity
(\ref{HSch}).

\subsection*{Acknowledgements}
It is my pleasure to dedicate this paper to Lothar Sch\"{a}fer on
the occassion of his $60^{th}$ birthday. The author is grateful to
Tom Spencer for early discussions on the derivation of the
nonlinear $\sigma-$ model outlined in the Appendix A, and to
Martin Zirnbauer for his interest in the work, constructive
criticism and many illuminating remarks. A useful communication
with Wei Yi who noticed inconsistencies in the derivation of
Appendix B of \cite{fystra2} is acknowledged. The work was
supported by EPSRC grant "Random Matrices and Polynomials: a tool
to understand complexity".
\section*{ Appendix A: From $k$-orbital model to
nonlinear $\sigma-$model without Hubbard-Stratonovich identity.}

Consider $N\times N$ random Hermitian matrix $\hat{H}$, with all
its entries $H_{lm}$ for $l\le m$ being independent random
Gaussian with the variance $\langle
H_{lm}^*H_{lm}\rangle=J_{lm}>0$. Let $N=rk$, with $r$ and $k$
being integers. Subdivide the index set $I=1,2,...,N$ into $r$
subsets $I_1,I_2,...,I_r$ each having exactly $k$ elements:
$I_i=\{(i-1)k+1,(i-1)k+2,...,(i-1)k+k\}$ and consider the
variances $J_{lm}$ such that
\begin{equation}\label{variances}
J_{lm}=\left\{\begin{array}{c} J/k\quad,\quad \mbox{if}\,\,( l,m\in I_i)\\
V/k^2\quad,\quad \mbox{if}\, (l\in I_i,m\in I_{i+1})\,\,\mbox{or}
\,\,(l\in I_{i-1}, m\in I_{i})\\ 0\quad
\mbox{otherwise}\end{array} \right.
\end{equation}
This defines the Wegner's gauge-invariant $k-$orbital
model\cite{SW} of $r$ "sites" arranged in the one-dimensional
lattice $d=1$.  Each "site" is a block $I_i$ representing a group
of $k$ orbitals, with $J/k$ standing for intragroup coupling and
$V/k^2$ representing the coupling of neighbouring groups. The
matrix $\hat{H}$ can be visualized as having a (block-)banded
structure made of $k\times k$ blocks, with nonzero entries
concentrated in a band of the widths $\propto k$ around the main
diagonal. The couplings  are scaled in a way ensuring correct
behaviour in the limit of infinitely many orbital $k\to\infty$,
which is assumed at a later stage (see also closely related model
in\cite{NVWY}). Generalization to lattices of higher spatial
dimensions is obvious.

Our goal here is to demonstrate that the method developed in
\cite{my2002} for random matrix problems without underlying
lattice structure
 can be straightforwardly applied to arrive to a
nonlinear $\sigma-$model from the $\hat{H}$ defined above. We
start with considering the simplest nontrivial object, the
negative integer moment of the modulus of the spectral determinant
of $\hat{H}$:
\begin{equation}
{\cal Z}_{n,n}(E,\eta)=
\left\langle|\det{\left(E+i\eta/k-\hat{H}\right)}|^{-2n}\right\rangle_H
\end{equation}
where $E$ and $\eta>0$ are real and brackets stand for the
ensemble averaging. Our first goal is to show that it has the
following integral representation:
\begin{eqnarray}\label{intrep1}
{\cal Z}_{n,n}(E,\eta)=Const \times \int d\hat{q}_1...\int
d\hat{q}_r e^{-k\sum_{i=1}^r{\cal L}(\hat{q}_i)}
\\ \nonumber
\times \prod_{i=1}^r
\frac{1}{\left[\det{\hat{q}_i}\right]^{2n}}\,\,
e^{-\frac{V}{2}\sum_{i=1}^{r-1} \mbox{\tiny
Tr}\left[\hat{q}_i\hat{L}\hat{q}_{i+1}\hat{L}\right]-
\eta\sum_{i=1}^{r}\mbox{\tiny Tr}\hat{q}_i}
\end{eqnarray}
where $ \hat{L}=\mbox{diag}({\bf 1}_n,-{\bf 1}_n)$
\begin{equation}
{\cal L}(\hat{q})=\frac{J}{2}\mbox{Tr}\hat{q}^2-iE\mbox{Tr}\hat{q}
-\mbox{Tr}\ln{\hat{q}}
\end{equation}
and the integration goes over the positive definite Hermitian
$2n\times 2n$ matrices $\hat{q}_i>0$. This representation is exact
(no approximations) with the only restriction $k\ge 2n$.

Here we outline the main steps employed to derive
Eq.(\ref{intrep1}). For every $l=1,...,N$ introduce two $n-$
component vectors $\Phi_{l,+}\,,\,\Phi_{l,-}$ with complex
components $S^{(1)}_{l,\pm}, ... ,S^{(n)}_{l,\pm}$ and also use
the notation
$\Psi_l=\left(\begin{array}{c} \Phi_{l,+}\\
\Phi_{l,-}\end{array}\right)$ and $\eta_k=\eta/k$. Representing
the inverse determinants as Gaussian integrals over $\Phi_{l,\pm}$
we have
\begin{eqnarray}\label{g}
{\cal Z}_{n,n}(E,\eta)\propto \int d\Psi_1...\int d \Psi_N
\exp\{\sum_{l=1}^N\Psi^{\dagger}_l \left(i E{ \hat{L}}-\eta_k{\bf
1}_{2n}\right)\Psi_l\}
\left\langle e^{-i\sum_{lm}H_{lm}\Psi^{\dagger}_l {
\hat{L}}\Psi_m}\right\rangle_H
\end{eqnarray}
Now for any $l=1,\ldots, N$ we introduce $2n\times 2n$ Hermitian
matrices $\hat{Q}_l=\Psi_l\otimes \Psi_l^{\dagger}$, so that the
result of the (gaussian) averaging for any variances $J_{lm}$ can
be written as
\[
\left\langle e^{-i\sum_{lm}H_{lm}\Psi^{\dagger}_l {
\hat{L}}\Psi_m}\right\rangle_H=e^{-\frac{1}{2}
\sum_{lm}J_{lm}\mbox{\small Tr}\left(\hat{Q}_l{ \hat{L}}
\hat{Q}_m{ \hat{L}}\right)}
\]

For the particular choice of the variances Eq.(\ref{variances}) we
can further write
\[
\sum_{lm}J_{lm}\mbox{Tr}\left(\hat{Q}_l{ \hat{L}} \hat{Q}_m{
\hat{L}}\right)=\frac{J}{k}\sum_{i=1}^r\mbox{Tr}
\left(\sum_{s=1}^k \hat{Q}_{s}^{(i)}{ \hat{L}}\right)^2
+\frac{V}{k^2}\sum_{i=1}^{r-1}\mbox{Tr} \left(\sum_{s=1}^k
\hat{Q}_{s}^{(i)}{ \hat{L}}\right) \left(\sum_{s=1}^k
\hat{Q}_{s}^{(i+1)}{ \hat{L}}\right)
\]
where we denoted $\hat{Q}_{s}^{(i)}\equiv \hat{Q}_{k(i-1)+s}$.
Introduce now the set of matrices $\hat{q}_i=\sum_{s=1}^k
\hat{Q}_{s}^{(i)}$ for $i=1,...,r$ and rewrite the integral as
\begin{eqnarray}\label{gg}
{\cal Z}_{n,n}(E,\eta)\propto \int d\hat{q}_1 ... d\hat{q}_r {\cal
I}(\hat{q}_1) ... {\cal I}(\hat{q}_r)
e^{-\frac{J}{2k}\mbox{Tr}\sum_{i=1}^r \left(\hat{q}_{i}{
\hat{L}}\right)^2 -\frac{V}{2k^2}\sum_{i=1}^{r-1}\mbox{Tr} \left(
\hat{q}_{i}{ \hat{L}}\right) \left(\hat{q}_{i+1}{ \hat{L}}\right)}
\end{eqnarray}
where
\begin{eqnarray}\label{I}
{\cal I}(\hat{q})=\int
d\Psi_1...d\Psi_k\delta\left(\hat{q}-\sum_{s=1}^k \Psi_s\otimes
\Psi_s^{\dagger}\right)\, \exp\left\{\sum_{s=1}^k\Psi^{\dagger}_s
\left(i E{ \hat{L}}-\eta_k{\bf 1}_{2n}\right)\Psi_s\right\}
\end{eqnarray}
Representing the $\delta-$function factor as the Fourier
transformation over the Hermitian $2n\times 2n$ matrix $\hat{K}$:
\[
\delta\left(\hat{q}-\sum_{s=1}^k \Psi_s\otimes
\Psi_s^{\dagger}\right)\propto \int d\hat{K} e^{i\mbox{\small
Tr}\hat{K}\hat{q}} e^{-i\sum_{s=1}^k\mbox{\small Tr}
\left[\hat{K}\left(\Psi_s\otimes \Psi_s^{\dagger}\right)\right]}
\]
and taking into account $\mbox{Tr}\left[\hat{K}\left(\Psi_s\otimes
\Psi_s^{\dagger}\right)\right]=\Psi_s^{\dagger}\hat{K}\Psi_s$ we
find that the $\Psi-$integrals are gaussian (and convergent in
view of $\eta_k>0$) and so when performed explicitly yield the
factor:
\[
\det{\left[\hat{K}-E{ \hat{L}}-i\eta_k{\bf 1}_{2n}\right]}^{-k}
\]
We immediately see that the resulting integral over $\hat{K}$ is
precisely one of Ingham-Siegel type calculated in \cite{my2002}
\footnote {A mathematically-minded reader who may dislike  a
little bit frivolous manipulations with $\delta-$functions, may
wish to perform the derivation by exploiting the "integration
theorem" proved in \cite{fystra1}. See also an alternative way of
derivation in \cite{rigor}.}. For $k\ge 2n$ it gives:
\begin{eqnarray}\label{II}
{\cal
I}(\hat{q})=\theta\left(\hat{q}\right)\left(\det{\hat{q}}\right)^{k-2n}
e^{i\mbox{\small Tr}\hat{q}\left(E{ \hat{L}}+i\eta_k{\bf
1}_{2n}\right)}
\end{eqnarray}
where the factor $\theta\left(\hat{q}\right)$ is non-zero for
positive definite matrices and zero otherwise. Substitute this
expression back to Eq.(\ref{gg}) and finally change $\hat{q}\to
k\,\,\hat{q}$ This immediately produces the formula
eq.(\ref{intrep1}).

Now change the integration variables in Eq.(\ref{intrep1}) from
$\hat{q}_i$ for $i=1,\ldots, r$ to the matrices
$\hat{\sigma}_i=\hat{q}_i\hat{L}$ parametrized as
$\hat{\sigma}_i=\hat{T}_i^{-1}\hat{P}_i\hat{T}_i$, with
 $\hat{T}_i\in U(n,n)/U(1)\otimes...\otimes U(1)$ and
$\hat{P}_i=\mbox{diag}(p^{(1)}_{1,i},...,
p^{(1)}_{n,i},p^{(2)}_{1,i},..., p^{(2)}_{n,i})$, such that
$p^{(1)}_{l,i}>0,p^{(2)}_{l,i}<0$. Correspondingly, for each
$i=1,...,r$ the integration measure is given by
$d\hat{\sigma}_i\propto \Delta^2(\hat{P}_i)
d\mu(T_i)d\hat{P}_{1,i}d\hat{P}_{2,i}$ and the above integral
assumes the form:
\begin{eqnarray}
{\cal Z}_{n,n}(E,\eta)=Const \times \int d\hat{\sigma}_1...\int
d\hat{\sigma}_r e^{-k\sum_{i=1}^r{\cal L}(\hat{P}_i)}
\\ \nonumber \times\prod_{i=1}^r \frac{1}{\left[\det{\hat{P}_i}\right]^{2n}}
\,\, e^{-\frac{V}{2}\sum_{i=1}^{r-1} \mbox{\tiny
Tr}\hat{\sigma}_i\hat{\sigma}_{i+1}- \eta\sum_{i=1}^{r}\mbox{\tiny
Tr}\hat{\sigma}_i\hat{L}}
\end{eqnarray}
where we denoted
\begin{equation}
{\cal L}(\hat{P})=\frac{J}{2}\mbox{Tr}\hat{P}^2-iE\mbox{Tr}\hat{P}
-\mbox{Tr}\ln{\hat{P}}
\end{equation}

Now it is evident that in the limit $k\to\infty$ all the matrices
$\hat{P}_i$ will be fixed by the (unique) saddle-point value
$\hat{P}_0$ minimizing the "action" ${\cal L}(\hat{P})$ and given
by
\begin{equation}\label{sp}
\hat{P}_0=\frac{1}{2J}\left(iE \,{\bf\hat{ 1}}_{2n}+
\hat{L}\,\sqrt{4J-E^2}\right)
\end{equation}
 At the next step one should
 accurately perform the
integration of Gaussian fluctuations around the saddle-point (
presence of the Vandermonde determinants makes the pre-exponential
factors vanishing at the saddle-point for any $n\ge 2$). The
resulting expression will clearly have the following structure:
\begin{equation}
{\cal Z}_{n,n}(E,\eta)\propto \int d{\cal Q}_1...\int d{\cal Q}_r
\{\ldots\} \, \,e^{-\frac{V}{2J}\sum_{i=1}^{r-1} \mbox{\small
Tr}{\cal Q}_i{\cal Q}_{i+1}-
\frac{\eta}{\sqrt{J}}\sum_{i=1}^{r}\mbox{\small Tr}{\cal
Q}_i\hat{L}}
\end{equation}
 where the integration manifold is parametrized by
the matrices ${\cal Q}_i=\hat{T}_i^{-1}\hat{L}\hat{T}_i$.  Here we
used dots to denote pre-exponential factors which may possibly
arise and also set the energy parameter $E=0$ for simplicity. This
is exactly the lattice version of the nonlinear $\sigma-$ model
introduced by Wegner and Sch\"{a}fer\cite{SW}. The index
$i=1,\ldots,r$ numbers the lattice sites and for $r\to \infty$ in
the lattices of dimensions $d>2$ the effective coupling constant
$\frac{V}{2J}$ controls the transition from localised to extended
states in the underlying Hamiltonian $H$.

 The saddle-point procedure at $k\to\infty$ and
subsequent manipulations are still to be done rigorous by strict
mathematical standards, see recent progress and discussions of
related issues in \cite{rigor,diser} .
 Another interesting problem is how to include anticommuting degrees
of freedom in the above derivation. Technically this can be done
following various methods, and will be discussed elsewhere
\cite{my}, see also \cite{Martin}.

\section*{Appendix B: proof of the formula (\ref{HSch8}).}

The starting point of the proof is the identity (\ref{HSchsimp}).
Introduce the singular value decompositions of the general complex
matrices $\hat{R},\hat{A}$ as:
\begin{equation}\label{1A}
\hat{R}=\hat{U}\hat{P}\hat{V}^{\dagger},\,\, \hat{P}=\mbox{diag}
(p_1,\ldots,p_n)>0,\quad
\hat{A}=\hat{U}_A\hat{a}\hat{V}_A^{\dagger},\,\,
\hat{a}=\mbox{diag} (a_1,\ldots,a_n)>0
\end{equation}
where $\hat{U},\hat{U}_A,\hat{V},\hat{V}_A$ are $n\times n$
unitary, with the associated Haar's measures $d\mu(\hat{U})$ and
$d\mu(\hat{V})$, respectively. The volume element ${\cal
D}\hat{R}{\cal D}\hat{R}^{\dagger}$ in new coordinates associated
with the singular value decomposition is given by ${\cal
D}\hat{R}{\cal D}\hat{R}^{\dagger}\propto \prod_{l=1}^{n}p_l
dp_l\prod_{l<m}(p_l^2-p_m^2)^2d\mu(\hat{U})\,d\mu(\hat{V})$.

Introducing $\tilde{U}=\hat{U}^{\dagger}\hat{U}_A$ and
$\tilde{V}^{\dagger}=\hat{V}_A^{\dagger}\hat{V}$, and using the
invariance of the Haar's measures:
$d\mu(\hat{U})=d\mu(\tilde{U})$, $d\mu(\hat{V})=d\mu(\tilde{V})$,
we rewrite the integral in the right-hand side of (\ref{HSchsimp})
as:
\begin{equation}\label{2A}
\int_{0}^{\infty} \prod_{l=1}^n dp_l p_l\,
\Delta^2[\hat{P}^2]\,e^{-\sum_{l=1}^n p_l^2}\int
d\mu(\tilde{U})\int d\mu(\tilde{V})e^{-i\mbox{\small Tr}
[\hat{P}(\tilde{U}\hat{a}\tilde{V}^{\dagger}+\tilde{V}\hat{a}\tilde{U}^{\dagger})]}
\end{equation}
The integral over two unitary matrices in this expression is
well-known from the papers by Guhr and Wettig \cite{guhr} and
Jackson et al.\cite{verb}:
\begin{equation}\label{TINTEGRAL1}
\int d\mu(\hat{U})d\mu(\hat{V}) e^{-i\mbox{\small
Tr}\left(\hat{P}\left[\hat{U}\hat{a}_d\hat{V}^{\dagger}+
\hat{V}\hat{a}_d\hat{U}^{\dagger} \right]\right)} \propto
\;\frac{\mbox{det} \left[J_0(2p_ia_j)\right]\vert_{1\leq i,j\leq
n}}{\triangle(p_1^2,\ldots ,p_n^2)\;\triangle(a_1^2,\cdots
,a_n^2)}
\end{equation}
Now substitute (\ref{TINTEGRAL1}) into (\ref{2A}), and after
cancelling one Vandermonde factor observe that due to
the invariance of the integrand all $n!$ terms of the determinant
made of Bessel functions yield identical contributions. Disregarding
the multiplicative constants
one can therefore replace that determinant with
$\prod_{i}^n J_0(2p_ia_i)$, so that
the right-hand side of (\ref{HSchsimp}) takes the form
\begin{equation}\label{4A}
\frac{1}{\Delta[\hat{a}^2]}
\int_{0}^{\infty} \prod_{l=1}^n dp_l p_l\,
\Delta[\hat{P}^2]\,e^{-\sum_{l=1}^n p_l^2}
\prod_{l}^n J_0(2p_la_l)
\end{equation}
On the other hand, the left hand side is $\exp{-\mbox{\small Tr}
\hat{A}^{\dagger}\hat{A}}\equiv e^{-\sum_{l=1}^n a_l^2}$ which
proves the formula (\ref{HSch8}).

\section*{Appendix C. Matrix Macdonald functions
associated with integrals over complex matrices}

The integral (\ref{TINTEGRAL1}) can be looked at as certain matrix
Bessel function that corresponds to Itzykson-Zuber-like integrals
over unitary matrices, see details in Guhr and Wettig, Guhr and
Kohler \cite{guhr,guhrk}. Below we consider the matrix Macdonald
functions that are associated with integrals over arbitrary
complex matrices:
\begin{equation}\label{TINTEGRAL2}
\int d\mu(\hat{T},\hat{T}^{\dagger})
e^{-\frac{1}{2}\mbox{Tr}\left(\hat{X}_d
\left[\hat{T}\hat{Y}_d\hat{T}^{\dag}+\left(\hat{T}^{\dag}\right)^{-1}
\hat{Y}_d\hat{T}^{-1}\right]\right)}
=\mbox{const}\;\frac{\mbox{det}\left[K_0(x_iy_j)\right]\vert_{1\leq
i,j\leq n}}{\triangle(x_1^2,\ldots ,x_n^2)\;\triangle(y_1^2,\cdots
,y_n^2)}
\end{equation}
where $\hat{X}_d$ and $\hat{Y}_d$ are two positive definite
diagonal matrices:
\begin{equation}
\hat{X}_d=\mbox{diag}\left(x_1,x_2,\ldots ,x_n\right)>0,\;\;
\hat{Y}_d=\mbox{diag}\left(y_1,y_2,\ldots ,y_n\right)>0
\end{equation}
The integration in (\ref{TINTEGRAL2}) goes over $\hat{T}\in
{\sf{Gl(n,{\mathcal{C}})/U(1)\times ... \times U(1)}}$, i.e.
complex matrices with real positive diagonal elements, and
$d\mu(\hat{T},\hat{T}^{\dagger})$ is the corresponding measure.
Note, however that the integrand is not changed if we consider
$\hat{T}\in {\sf{Gl(n,{\mathcal{C}})}}$ rather than restricting it
to the coset space as above. Therefore, if we use the invariant
Haar's measure $d\mu_H(\hat{T})$ for the full group, the result of
integration will be the same up to a constant factor.

  Let us look at the integral in (\ref{TINTEGRAL2}) as a
function $ \Phi(X_d,Y_d)$. We start with considering a pair of
$n\times n$ Hermitian positive definite matrices $\hat{X}^{(1)},
\hat{X}^{(2)} $, and another such pair $\hat{A},\hat{B}$.
Introduce the Laplace operator $D_{X^{(1)},X^{(2)}}$ acting on
such matrices as:
\begin{equation}
D_{X^{(1)},X^{(2)}}=\frac{1}{2} \sum\limits_{i\leq i,j\leq
n}\frac{\partial^2}{\partial(\mbox{Re}X^{(1)}_{ij})
\partial(\mbox{Im}X^{(2)}_{ji})}
\end{equation}
Then we  construct a function $W(X^{(1)},X^{(2)},A,B)$ with the property
\begin{equation}\label{laplacematrixequation1}
D_{X^{(1)},X^{(2)}}\,W(X^{(1)},X^{(2)},A,B)=
\mbox{Tr}(AB)\,W(X^{(1)},X^{(2)},A,B)
\end{equation}
In particular, the following function
\begin{equation}
W(X^{(1)},X^{(2)},A,B)=\exp\left[-\frac{1}{2}\mbox{Tr}\left(X^{(1)}A+
BX^{(2)}\right)\right]
\end{equation}
satisfies the Eq.(\ref{laplacematrixequation1}) as can be checked
by direct calculations. Let us now use a possibility
\cite{fystra2} to parametrize $A=T_Y Y_d T_Y^{\dag}$ and
$B=(T_Y^{\dag})^{-1}Y_d T_Y^{-1}$ in
Eq.(\ref{laplacematrixequation1}), as well as $\hat{X}^{(1)}=T_X
X_d T^{\dagger}_X, \hat{X}^{(2)}=(T^{\dagger}_X)^{-1} X_d
(T_X)^{-1}$. Here both $T_X,T_Y$ belong to the coset space
${\sf{Gl(n,{\mathcal{C}})/U(1)\times ... \times U(1)}}$.

We obtain:
\begin{eqnarray}\label{prec}
D_{X^{(1)},X^{(2)}}\exp{-\frac{1}{2}\mbox{Tr}\left(
\left[X^{(1)}T_YY_d T_Y^{\dag}+(T_Y^{\dag})^{-1}Y_d
T_Y^{-1}X^{(2)}
\right]\right)}=\nonumber\;\;\;\;\;\;\;\;\;\;\;\\
\mbox{Tr}(Y_d^2)\exp\left[-\frac{1}{2}\mbox{Tr}\left(X_d\left[\tilde{T}
Y_d\tilde{T}^{\dag}+(\tilde{T}^{\dag})^{-1}Y_d\tilde{T}^{-1}\right]\right)\right]
\end{eqnarray}
where we introduced the matrix $\tilde{T}=T_X^{\dagger}T_Y$
belonging to the group ${\sf{Gl(n,{\mathcal{C}})}}$.

 Now we use that (i) the integration over complex
matrices $T_Y$ commutes with the Laplace operator
$D_{X^{(1)},X^{(2)}}$ and (ii) for any fixed $T_X$ the result of
integrating the right hand side of (\ref{prec}) over $T_Y$ does
not depend on $T_X$ due to the possibility to use the invariant
measure $d\mu_H(T_Y)=d\mu_H(\tilde{T})$. We therefore conclude
that the matrix function $\Phi(X_d,Y_d)$ defined by the integral
Eq.(\ref{TINTEGRAL2}) satisfies the following differential
equation:
\begin{equation}\label{MATRIXBESSEL}
D_{X^{(1)},X^{(2)}}\Phi(X_d,Y_d)=\mbox{Tr}(Y_d^2)\Phi(X_d,Y_d)
\end{equation}
To derive the explicit formula Eq.(\ref{TINTEGRAL2}) for the
matrix function $\Phi(X_d,Y_d)$ we apply the method proposed by
Guhr and Wettig \cite{guhr} duly modified. We notice that when
passing from $X^{(1)},X^{(2)}$ to the "angular" coordinates $T_X$
and "radial" coordinates $X_d$ the radial part of the Jacobian is
given by $J(X_d)=\triangle^2(x_1^2,\ldots
,x_n^2)\prod\limits_{i=1}^{n}x_i$. This is enough to ensure that
the radial part $D_{X_d}$ of the Laplace operator
$D_{X^{(1)},X^{(2)}}$ must have the following expression:
\begin{equation}
D_{X_d}=\frac{1}{J(x)}\sum\limits_{i=1}^{n}\partial_iJ(x)\partial_i,\;
J(x)=\triangle^2(x_1^2,\ldots
,x_n^2)\prod\limits_{i=1}^{n}x_i
\end{equation}
Guhr and Wettig noted \cite{guhr} that the radial part $D_{X_d}$
of such form is, in fact, separable. It means that for an
arbitrary function $f(x_1,x_2,\ldots ,x_n)$ the following identity
holds:
\begin{equation}\label{separability}
D_{X_d}\frac{f(x_1,\ldots ,x_n)}{\triangle(x_1^2,\ldots
,x_n^2)}=\frac{1}{\triangle(x_1^2,\ldots
,x_n^2)}\sum\limits_{k=1}^{n}\left(\frac{\partial^2}{\partial
x^2_k}+\frac{1}{x_k}\frac{\partial}{\partial
x_k}\right)\frac{f(x_1,\ldots ,x_n)}{\triangle(x_1^2,\ldots
,x_n^2)}
\end{equation}
The separability of the operator $D_{X_d}$ enables us to solve the
differential equation Eq.(\ref{MATRIXBESSEL}) and to prove formula
(\ref{TINTEGRAL2}). The remaining parts of the proof given in the
appendix B of \cite{fystra2} hold its validity without any
modifications.


\begin{thebibliography}{99}
\bibitem{Efetov} K.B.Efetov {\it Adv.Phys.} {\bf 32},53 (1983);
K.B.Efetov "Supersymmetry in Disorder and Chaos"
Cambridge University Press, Cambridge 1997
\bibitem{Mirlin} A.D.Mirlin {\it Physics Reports} {\bf 36} (2000)
259
\bibitem{Wegner} F.Wegner {\it Zeitsch. Physik} {\bf 36} (1979), 207
\bibitem{SW} L. Sch\"{a}fer, F. Wegner {\it Z.Phys.B} {\bf 38} (1980)
113
\bibitem{PS} A.M.M. Pruisken and L.Sch\"{a}fer {\it Nucl.Phys.B}
{\bf 200} (1982), 20
\bibitem{rigor} T.Spencer and M.R Zirnbauer "Spontaneous symmetry
breaking of a hyperbolic sigma-model in three dimensions",
preprint 2004
\bibitem{VWZ} J.J.M. Verbaarschot, H.A.Weidenm\"{u}ller,
M.R.Zirnbauer {\it Phys.Rep.} {\bf 129}, 367 (1985)
\bibitem{FySo} Y.V.Fyodorov and H.-J.Sommers {\it J.Math.Phys.}
{\bf 38}, 1918 (1997)
\bibitem{VW} J.J.M. Verbaarschot and T.Wettig {\it
Annu.Rev.Nucl.Part.Sci.} {\bf 50} 343 (2000)
\bibitem{my95} Y.V.Fyodorov in: E.Akkemans et al. (Eds.)
Mesocopic Quantum Physics, Les Houches, Session LXI, 1994,
Elsevier, Amsterdam, 1995; p.493
\bibitem{Martin} M.R. Zirnbauer "The supersymmetry method of
random matrix theory", {\it e-print } math-ph/0404057
\bibitem{my2002} Y.V. Fyodorov {\it Nucl.Phys.B} {\bf 621} [PM] (2002)
643-674
\bibitem{band} Y.V. Fyodorov and A.D.Mirlin {\it Phys.Rev. Lett.}
{\bf 67} (1991) 2405 and {\it Int.J.Mod.Phys. B} {\bf 8} (1994) 3795
\bibitem{Martin0} M.R. Zirnbauer {\it J.
Math. Phys.} {\bf 37} (1996) 4986; see the end of Section III.E,
on page 5006
\bibitem{fystra1} Y.V. Fyodorov and E Strahov {\it Nucl.Phys.B}
{\bf 630} [PM] (2002) 453
\bibitem{RG} I.S.Gradshteyn and I.M.Ryzyk "Table of Integrals,Series and Products"
(ed. A.Jeffrey), 6th ed. (Academic Press (San Diego), 2000); (a)
formula 8.421 and definitions 8.405; (b) formula 6.631.4
\bibitem{Martin1} M.R. Zirnbauer "Symmetry classes in Random Matrix
Theory" , {\it e-print } math-ph/0404058
\bibitem{fystra2}Y.V. Fyodorov and E Strahov {\it Nucl.Phys.B}
{\bf 647} [PM] (2002) 581
\bibitem{IZHC} C.Itzykson, J.B.Zuber {\it J. Math.Phys.}{\bf 21}
(1980) 411; Harish-Chandra {\it Proc.Nat.Acad.Sci.}{\bf 42} (1956) 252
\bibitem{guhr} T. Guhr and T.Wettig {\it J.Math.Phys.}{bf 37}
(1996) 6395; F.A.Berezin and F.I.Karplevich {\it Dokl.Akad.Nauk.SSSR}
{\bf 118} (1958), 9
\bibitem{verb} A.D.Jackson, M.K.Sener and J.J.M.Verbaarschot
{\it Phys.Lett.B} {\bf 387} (1996) 355
\bibitem{guhrk} T.Guhr and H.Kohler {\it J.Math.Phys. } {\bf 43}
(2002)2707
\bibitem{NVWY} N. Nishioka, J.J.M Verbaarschot, H.A.Weidenm\"{u}ller
and S. Yoshida {\it Ann. Phys} {\bf 172} (1986) 67-99
\bibitem{diser} M.Disertori, H.Pinson and T.Spencer
{\it Commun.Math.Phys.} {\bf 232}, (2002), 83
\bibitem{my} Y.V. Fyodorov, unpublished

\end{thebibliography}
\end{document}